\documentclass[12pt,letterpaper]{article}

\usepackage{amsmath,amssymb,calc}
\usepackage{graphicx}

\newcommand\be{\begin{equation}}
\newcommand\ee{\end{equation}}
\newcommand{\bea}{\begin{eqnarray}}
\newcommand{\eea}{\end{eqnarray}}

\newcommand{\nn}{\nonumber}
\newcommand{\pd}{\partial}

\def\id{\protect{{1 \kern-.28em {\rm l}}}}

\def\id{\protect{{1 \kern-.28em {\rm l}}}}

\let\non\nonumber

\begin{document}

\begin{titlepage}
\begin{center}
\hfill \\
\vspace{2cm}
{\Large {\bf De Sitter Space in Gauge/Gravity Duality
\\[3mm] }}

\vskip 1.3cm
{\bf Lilia Anguelova${}^a$\footnote{anguelova@inrne.bas.bg}, Peter Suranyi${}^b$\footnote{peter.suranyi@gmail.com} and L.C.R. Wijewardhana${}^b$\footnote{rohana.wijewardhana@gmail.com}\\
\vskip 0.5cm  {\it ${}^a$ Institute for Nuclear Research and Nuclear Energy, BAS, Sofia, Bulgaria\\ ${}^b$ Department of Physics, University of Cincinnati,
Cincinnati, OH 45221, USA}\non\\}


\vskip 6mm

\end{center}

\vskip .1in
\vspace{0.6cm}

\begin{center} {\bf Abstract}\end{center}

\vspace{-1cm}

\begin{quotation}\noindent

We investigate gauge/gravity duality for gauge theories in de Sitter space. More precisely, we study a five-dimensional consistent truncation of type IIB supergravity, which encompasses a wide variety of gravity duals of strongly coupled gauge theories, including the Maldacena-Nunez solution and its walking deformations. We find several solutions of the 5d theory with $dS_4$ spacetime and nontrivial profiles for (some of) the scalars along the fifth (radial) direction. In the process, we prove that one of the equations of motion becomes dependent on the others, for nontrivial warp factor. This dependence reduces the number of field equations and, thus, turns out to be crucial for the existence of solutions with $(A)dS_4$ spacetime. Finally, we comment on the implications of our $dS_4$ solutions for building gravity duals of Glueball Inflation. 

\end{quotation}
\vfill
\flushleft{\today}

\end{titlepage}

\eject

\tableofcontents

\section{Introduction}

Since the astronomical observations of \cite{CosmAccel} indicated that the expansion of the Universe is accelerating at present, which is consistent with having a (small, but non-vanishing) positive cosmological constant, there has been a huge interest in understanding quantum field theory in de Sitter space \cite{QFTdS}, as well as in finding dS solutions in string compactifications with stabilized moduli \cite{dSST}. However, despite recent progress in either topic, these are still rather difficult endeavors. This has motivated a search for alternative approaches in the vein of attempts to develop holographic descriptions of physics in de Sitter space via proposed dS/CFT \cite{AS} or dS/dS \cite{dSdS} correspondences.\footnote{In this ``dS/dS" correspondence the relation is between a $dS_d$ space and a CFT on a $dS_{d-1}$ space.}

The most promising direction, though, for studying nonperturbative effects in gauge theories living in de Sitter space is via extensions of the gauge/gravity duality, that arose from \cite{JM,GKP,EW,IMSY}, with a curved 4d spacetime instead of a flat one. This kind of idea was explored in \cite{GIN}, where a solution of type IIB supergravity with a Wick rotated RR scalar was used.\footnote{See also \cite{FKP} for a holographic model of QFTs in de Sitter space, based on a $dS_d$ foliation of a $(d+1)$-dimensional asymptotically $AdS$ space. This model was further used in \cite{FNPT} to study thermal QFT properties, as well as the Schwinger effect in de Sitter space. Also worth mentioning are the works \cite{AB}, in which the gauge/gravity duality is used to try to build time-dependent backgrounds in supergravity/string theory; despite not having explicit  solutions, the authors manage to extract some interesting information by analyzing the UV and IR asymptotic behaviors of certain $dS$-deformed equations of motion.} The ten-dimensional metric of that solution has a 4d de Sitter part, that can be viewed as the spacetime for a certain gauge theory. The same truncation of type IIB, with Wick rotated RR scalar, was also utilized in \cite{EGM} to study more involved 4d cosmological backgrounds. However, this solution is rather peculiar, due to its reliance on a Wick rotated 10d field.

Our goal here will be to find solutions with a $dS_4$ spacetime within the framework of the five-dimensional consistent truncation of type IIB, established in \cite{BHM}, that describes the gravity duals of a broad class of confining gauge theories. In particular, this framework encompasses the Maldacena-Nunez \cite{MN} and Klebanov-Strassler \cite{KS} solutions, which provide gravitational duals to ${\cal N}=1$ SYM, as well as the more recently found solutions \cite{NPP,ENP,EGNP}, which give duals to strongly coupled gauge theories with more than one dynamical scale.

We will study the equations of motion of the effective 5d action relevant for the consistent truncation of \cite{BHM}. We will show that, for a metric ansatz compatible with a 4d cosmological constant (of either sign), there is a reduction in the number of independent equations by one, under a certain condition. This is of crucial importance for the existence of solutions, as otherwise generically the system is overdetermined for the physically desirable metric ansatz. We establish this result for a cosmological constant of either sign, because $AdS_4$ solutions in this context are in principle also of interest, for example, for realizations of the Karch-Randall model of locally localized gravity \cite{KR}. We will not investigate $AdS_4$ solutions here, though.

We will find several solutions with $dS_4$ spacetime and with only a subset of the 5d fields having nontrivial profiles along the fifth (radial) direction. These solutions occur within certain approximations, similar to the approximation in which the walking solution of \cite{NPP} is found. It would be interesting to understand, in the future, whether that similarity has a deeper meaning. Finally, we will discuss the relevance of the present considerations for Cosmological Inflation. Of course, de Sitter space is the leading approximation to the spacetime during Inflation. But, more importantly, being able to address strongly-coupled gauge dynamics during Inflation is of great significance for composite models \cite{CJS,BCJS}, in which the role of the inflaton is played by a quark condensate or a glueball. Our solutions with a $dS_4$ spacetime could provide the starting point for building a gravity dual of Glueball Inflation, since a slowly varying Hubble parameter (as is the case in slow roll Inflation) can be viewed as a small time-dependent deformation around a constant one (the latter corresponding to the pure de Sitter space). 

In the next section, we review relevant material about the consistent truncation of \cite{BHM}. In Section 3, we explain our ansatz for the 5d fields, including the metric, and derive the resulting equations of motion. Furthermore, we show that, under a certain condition, one of those equations becomes dependent on the others and, as a result, is automatically satisfied when they are. We also show that there is an additional consistent truncation in the scalar sector that one can make. In Section 4, we look for solutions of the equations of motion. In Subsection 4.1 we find analytically an explicit and rather simple solution with a positive 4d cosmological constant. In Subsection 4.2 we show numerically that there is an interesting class of solutions with a natural upper bound for the radial variable. In Appendix A we investigate the allowed parameter space for those numerical solutions. Finally, in Sections 5 and 6 we discuss the relevance of our considerations for Glueball Inflation and summarize the results of this paper.

\section{Consistent truncation to a 5d theory}
\setcounter{equation}{0}

We will look for solutions within the consistent truncation of type IIB to a 5d theory, found in \cite{BHM}. This set-up encompasses a variety of gravity duals of confining strongly-coupled gauge theories including the famous Maldacena-Nunez \cite{MN} and Klebanov-Strassler \cite{KS} ${\cal N} = 1$ solutions, as well as their deformations \cite{NPP,ENP,EGNP} describing gauge theories with multi-scale dynamics. 

The bosonic fields of IIB supergravity are the 10d metric $g_{AB} (x^M)$, the string dilaton $\phi (x^M)$, the RR scalar $C (x^M)$, the NS  3-form field-strength $H_3 (x^M)$, and the RR 3-form $F_3 (x^M)$ and 5-form $F_5 (x^M)$ fields, where $A,B,M = 0,...,9$. The ansatz for the consistent truncation of interest for us is the following. The metric is:
\bea \label{10dmetric}
ds_{10d}^2 &=& e^{2p-x} ds_{5d}^2 \,+ \,e^{x+g} (\omega_1^2 \,+ \,\omega_2^2) \,+ \,e^{x-g} \!\left[ (\tilde{\omega}_1 + a \omega_1)^2 \!+ \!(\tilde{\omega}_2 - a \omega_2)^2 \right] \nn \\
&+& \, e^{-6p-x} (\tilde{\omega}_3 + \omega_3)^2 \qquad , \qquad s_{5d}^2 = g_{IJ} \,dx^I dx^J \qquad ,
\eea
where
\bea \label{omegatilde}
\tilde{\omega}_1 &=& \cos \psi d\tilde{\theta} + \sin \psi \sin \tilde{\theta} d \tilde{\varphi} \,\, , \hspace*{2cm} \omega_1 = d \theta \,\, , \nn \\
\tilde{\omega}_2 &=& - \sin \psi d\tilde{\theta} + \cos \psi \sin \tilde{\theta} d \tilde{\varphi} \,\, , \hspace*{1.6cm} \omega_2 = \sin \theta  d \varphi \,\, , \nn \\
\tilde{\omega}_3 &=& d \psi + \cos \tilde{\theta} d \tilde{\varphi} \,\, , \hspace*{3.7cm} \omega_3 = \cos \theta d \varphi \,\, .
\eea
The RR 3-form is:
\bea
F_3 &=& P \left[ - (\tilde{\omega}_1 + b \,\omega_1) \wedge (\tilde{\omega}_2 - b \,\omega_2) \wedge (\tilde{\omega}_3 + \omega_3) \right. \nn \\ 
&+& \left. \,(\pd_I b) \,dx^I \!\wedge (- \omega_1 \wedge \tilde{\omega}_1 + \omega_2 \wedge \tilde{\omega}_2) + (1-b^2) (\omega_1 \wedge \omega_2 \wedge \tilde{\omega}_3) \right] ,
\eea
where $P=const$. The remaining fields are:
\be
\phi = \phi (x^I) \qquad , \qquad C = 0 \qquad , \qquad H_3 = 0
\ee
and
\be \label{F5}
F_5 = {\cal F}_5 + \star {\cal F}_5 \qquad , \qquad {\cal F}_5 = Q \,vol_{5d} \qquad , \qquad Q=const \quad .
\ee
Finally, the quantities $p$, $x$, $g$, $a$, $b$, $\phi$ in the above ansatz are all functions of the 5d coordinates $x^I$. In other words, these are six scalars in the 5d external space. Similarly, the 5d metric $g_{IJ}$ in (\ref{10dmetric}) also depends on $x^I$, i.e. $g_{IJ} = g_{IJ} (x^I)$.

Note that the full consistent truncation of \cite{BHM} allows for $H_3 \neq 0$ and that is, in fact, needed to obtain the Klebanov-Strassler solution \cite{KS}; see \cite{BHM,PT}. However, it is consistent to set the NS three form $H_3 = 0$ and we will do so in the following for simplicity.\footnote{We will comment more on the consistency of taking $H_3=0$ later.} This smaller consistent truncation still encompasses the Maldacena-Nunez solution \cite{MN} and its deformations \cite{NPP,ENP}, which describe walking gauge theories.

Substituting (\ref{10dmetric})-(\ref{F5}) into the ten-dimensional IIB action and integrating out the compact internal dimensions, parameterized by the angular coordinates $\theta$, $\varphi$, $\tilde{\theta}$, $\tilde{\varphi}$ and $\psi$, one finds that the five-dimensional fields $\Phi^i (x^I) = \{ \,p  (x^I), x (x^I), g (x^I), \phi (x^I), a (x^I), b (x^I) \,\}$ and $g_{IJ} (x^I)$ are described by the action:
\be \label{S5d}
S = \int d^5 x \sqrt{- det g} \left[ - \frac{R}{4} + \frac{1}{2} \,G_{ij} (\Phi) \,\pd_{I} \Phi^i \pd^{I} \Phi^i + V (\Phi) \right] \, ,
\ee
where the sigma model metric $G_{ij} (\Phi)$ is diagonal and has the components
\be \label{SigmaMM}
G_{pp} = 6 \,\,\, , \,\,\, G_{xx} = 1 \,\,\, , \,\,\, G_{gg} = \frac{1}{2} \,\,\, , \,\,\, G_{\phi \phi} = \frac{1}{4} \,\,\, , \,\,\, G_{aa} = \frac{e^{-2g}}{2} \,\,\, , \,\,\, G_{bb} = \frac{P^2 e^{\phi - 2x}}{2}
\ee
and, finally, the potential $V (\Phi)$ has the form
\bea \label{Pot}
V (\Phi) &=& - \,\frac{e^{2p - 2x}}{2} \!\left[ e^g + (1+a^2) e^{-g} \right] + \frac{e^{-4p-4x}}{8} \!\left[ e^{2g} +(a^2-1)^2 e^{-2g} + 2a^2 \right] \nn \\
&+& \frac{a^2}{4} \,e^{-2g+8p} + P^2 \,\frac{e^{\phi-2x+8p}}{8} \!\left[ e^{2g} + e^{-2g} (a^2-2ab+1)^2 + 2(a-b)^2 \right] \nn \\
&+& Q^2 \,\frac{e^{8p-4x}}{8} \,\,\, .
\eea

Let us note that the equations of motion, that follow from the action (\ref{S5d}), are:
\bea \label{EoM}
\nabla^2 \Phi^i + {\cal G}^i{}_{jk} \,g^{IJ} (\pd_I \Phi^j) (\pd_J \Phi^k) - V^i &=& 0 \quad , \nn \\
- R_{IJ} + 2 \,G_{ij} \,(\pd_I \Phi^i) (\pd_J \Phi^j) + \frac{4}{3} \,g_{IJ} V &=& 0 \quad ,
\eea
where $\nabla^2 = \nabla_I \nabla^I$ and $V^i = G^{ij} V_j$ with $V_j \equiv \frac{\pd V}{\pd \Phi^j}$. From (\ref{SigmaMM}) it is easy to compute that the Christoffel symbols ${\cal G}^i{}_{jk}$ for the metric $G_{ij}$ are:
\bea \label{SigmaMCh}
&&{\cal G}^b_{\phi b} = \frac{1}{2} \quad , \quad {\cal G}^b_{x b} = -1 \quad , \quad {\cal G}^{a}_{g a} = - 1 \quad , \quad {\cal G}^g_{aa} = e^{-2g} \nn \\
&&{\cal G}^{\phi}_{bb} = - P^2 e^{\phi - 2 x} \quad , \quad {\cal G}^x_{bb} = \frac{P^2 e^{\phi - 2x}}{2}
\eea
with all other components vanishing.

\section{Metric ansatz and field equations}
\setcounter{equation}{0}

In the context of the gauge/gravity duality, one can find gravitational duals of some strongly coupled gauge theories living in Minkowski space by solving (\ref{EoM}) with the metric ansatz
\be
ds_5^2 = e^{2A(z)} \eta_{\mu \nu} dx^{\mu} dx^{\nu} + dz^2 \,\, ;
\ee
the scalars $\Phi^I$ then describe glueball states in the dual gauge theory. For example, the Maldacena-Nunez solution \cite{MN} is obtained for 
\be
Q=0 \quad , \quad b=a \quad , \quad x=\frac{1}{2} g - 3p \quad , \quad \phi = -6p - g - 2 \ln P
\ee
and some particular functions $a=a(z)$, $g=g(z)$ and $p=p(z)$; for more details see \cite{BHM,PT}.\footnote{Note that here by definition $z$ is a radial coordinate running in the range $(0,\infty)$, often denoted in the literature by $\rho$ or $r$, {\it not} to be confused with a worldvolume radial coordinate running in the range $(-\infty,\infty)$ and relevant when studying U-shaped probe brane embeddings. In the ten-dimensional up-lift given by (\ref{10dmetric}), $z$ is the usual radial direction moving away from the D-brane sources of the background.}

Here, instead, we will be interested in solutions of (\ref{EoM}) with a curved four-dimensional spacetime. More precisely, we will take the 5d metric ansatz to be:
\be \label{5dMetr}
ds_5^2 = e^{2A(z)} g_{\mu \nu} dx^{\mu} dx^{\nu} + dz^2 \,\, ,
\ee
where the 4d metric $g_{\mu \nu}$ is de Sitter or anti-de Sitter and thus satisfies 
\be \label{4dR}
R_{\mu \nu} = \Lambda g_{\mu \nu} \qquad {\rm with} \qquad \Lambda = const \,\, .
\ee 
Clearly, $dS_4$ corresponds to $\Lambda>0$, whereas $AdS_4$ to $\Lambda < 0$. The $dS$ case is of great interest because of both the Inflationary epoch in the Early Universe and the present day accelerated expansion of the Universe. The $AdS$ case is relevant for the Karch-Randall model of locally localized gravity \cite{KR} and the related study of defect CFTs \cite{dCFTs}. 

To be more explicit, let us write the $dS_4$ metric as:
\be \label{dSm}
g_{\mu \nu} dx^{\mu} dx^{\nu} = -dt^2 + e^{2 \sqrt{\frac{\Lambda}{3}} \,t} d\vec{x}^2
\ee
and the $AdS_4$ one as: 
\be \label{AdSm}
g_{\mu \nu} dx^{\mu} dx^{\nu} = e^{-2 \sqrt{-\frac{\Lambda}{3}} \,x_3} (-dt^2 + dx_1^2 + dx_2^2) +dx_3^2 \,\,\, .
\ee
With these metric ansatze, we will look for solutions of (\ref{EoM}) for some nontrivial scalar profiles of the form
\be 
\Phi^i = \Phi^i (z) \,\, .
\ee 

It is easy to realize that for both cases, (\ref{dSm}) and (\ref{AdSm}), the action of the 5d operator $\nabla^2 = \pd_I \pd^I + \Gamma^I_{IJ} \pd_J$ has the same form:
\be \label{Nab2}
\nabla^2 \Phi^i = (\Phi^i)'' + 4 A' (\Phi^i)' \,\, ,
\ee
where we have denoted $' \equiv \pd_z$. The reason is that for every value of $\mu = 0,...,3$ the 5d Christoffel symbol components $\Gamma^{\mu}_{\mu z} = A'$, as well as $\Gamma^z_{zz} = 0$, for {\it both} (\ref{dSm}) and (\ref{AdSm}). Hence the scalar field equations in (\ref{EoM}) are
\be
(\Phi^i)'' + 4 A' (\Phi^i)' + {\cal G}^i{}_{jk} (\Phi^j)' (\Phi^k)' - G^{ij} V_j = 0 
\ee
regardless of whether the 4d metric $g_{\mu \nu}$ in (\ref{5dMetr}) is taken to be de Sitter or anti-de Sitter. 

To write out explicitly the metric equations in (\ref{EoM}), let us first compute the components of the Ricci tensor. For the $dS$ case, namely (\ref{5dMetr}) with (\ref{dSm}) substituted, we easily find: 
\bea \label{RiccidS}
R_{tt} &=& - \Lambda + e^{2A} \left( 4 A'^2 + A'' \right) \nn \\
R_{x^n x^n} &=& e^{2 \sqrt{\frac{\Lambda}{3}} \,t} \left[ \Lambda - e^{2A} \left( 4A'^2 + A''^2 \right) \right] \quad , \quad n = 1,2,3 \nn \\
R_{zz} &=& - 4 A'' - 4 A'^2
\eea
with all other components vanishing. Using (\ref{RiccidS}), it is easy to see that the $(tt)$ component of the second equation in (\ref{EoM}) gives, up to an overall sign, exactly the same as the $(x^n x^n)$ components. That equation is:
\be \label{R11Eq}
- \Lambda e^{-2A} + 4A'^2 + A'' + \frac{4}{3} V = 0 \,\, .
\ee
Finally, the $(zz)$ component of (\ref{EoM}) gives:
\be \label{RzzEq}
4 A'' + 4 A'^2 + 2 G_{ij} (\Phi^i)' (\Phi^j)' + \frac{4}{3} V = 0 \,\, .
\ee
For the $AdS$ case, i.e. (\ref{5dMetr}) with (\ref{AdSm}) substituted, the Ricci tensor components are somewhat different:
\bea \label{RicciAdS}
R_{tt} &=& e^{-2 \sqrt{-\frac{\Lambda}{3}} \,x_3} \left[ - \Lambda + e^{2A} \left( 4 A'^2 + A'' \right) \right] \nn \\
R_{x^n x^n} &=& e^{-2 \sqrt{-\frac{\Lambda}{3}} \,x_3} \left[ \Lambda - e^{2A} \left( 4A'^2 + A''^2 \right) \right] \quad , \quad n = 1,2 \nn \\
R_{x^3 x^3} &=& \Lambda - e^{2A} (4A'^2 + A''^2) \nn \\
R_{zz} &=& - 4 A'' - 4 A'^2
\eea
However, taking into account the difference in the 4d metric, this leads again exactly to equations (\ref{R11Eq}) and (\ref{RzzEq}). 

Summarizing, the system of field equations that we want to study is:
\bea \label{EoMsyst}
(\Phi^i)'' + 4 A' (\Phi^i)' + {\cal G}^i{}_{jk} (\Phi^j)' (\Phi^k)' - G^{ij} V_j &=& 0 \nn \\
- \Lambda e^{-2A} + 4A'^2 + A'' + \frac{4}{3} V &=& 0 \nn \\
4 A'' + 4 A'^2 + 2 G_{ij} (\Phi^i)' (\Phi^j)' + \frac{4}{3} V &=& 0 \,\, .
\eea
The difference in (\ref{EoMsyst}), between having de Sitter and anti-de Sitter $g_{\mu \nu}$ metrics in (\ref{5dMetr}), is only in the sign of the 4d cosmological constant $\Lambda$. Note that there is one more equation in (\ref{EoMsyst}) than there are unknown functions $\Phi^i (z)$, $A(z)$. So, at first sight, it is not clear whether this system can have any solutions. However, we will show now that, in fact, one of the equations is not independent of the others. Namely, it is automatically satisfied whenever the rest of the equations of motion are solved.

\subsection{Dependent equation of motion} \label{DepEq}

Here we will show that one of the equations of motion in (\ref{EoMsyst}) is dependent on the remaining ones. The dependent equation turns out to be one of the last two, namely one of the metric field equations.

For convenience, let us introduce the following notation:
\bea \label{E123}
E1 &:& \quad (\Phi^i)'' + 4 A' (\Phi^i)' + {\cal G}^i{}_{jk} (\Phi^j)' (\Phi^k)' - G^{ij} V_j = 0 \nn \\
E2 &:& \quad - \Lambda e^{-2A} + 4 A'^2 + A'' + \frac{4}{3} V = 0 \nn \\
E3 &:& \quad 4 A'' + 4 A'^2 + 2 G_{ij} (\Phi^i)' (\Phi^j)' + \frac{4}{3} V = 0 \,\, .
\eea 
Now let us take combinations of $E2$ and $E3$ in such a way that $A''$ and $A'^2$ will appear in separate equations. Namely, consider:
\bea \label{N2N3}
N2 \equiv E3 - E2  &:& \qquad \Lambda e^{-2A} + 3 A'' + 2 G_{ij} (\Phi^i)' (\Phi^j)' = 0 \,\, , \\
N3 \equiv 4 E2 - E3 &:& \qquad -4 \Lambda e^{-2A} + 12 A'^2 + 4 V - 2 G_{ij} (\Phi^i)' (\Phi^j)' = 0 \,\, . \nn
\eea 
From $N2$ we have:
\be \label{Adp}
A'' = - \frac{1}{3} \Lambda e^{-2A} - \frac{2}{3} G_{ij} (\Phi^i)' (\Phi^j)' \,\, .
\ee
Note that, for $\Lambda > 0$, the above relation implies the condition $A'' < 0$ for there to be a solution, since $G_{ij}$ is diagonal and with positive components; see (\ref{SigmaMM}). 

Next, let us differentiate the left hand side of $N3$ with respect to $z$:
\be \label{LdN3}
L_{dN3} \equiv 8 \Lambda A' e^{-2A} + 24 A' A'' + 4 V_i (\Phi^i)' - 2 \pd_k G_{ij} (\Phi^k)' (\Phi^i)' (\Phi^j)' - 4 G_{ij} (\Phi^i)'' (\Phi^j)' \, . 
\ee
Substituting (\ref{Adp}) into (\ref{LdN3}), we find:
\be \label{LdN3p}
L_{dN3} = - 16 A' G_{ij} (\Phi^i)' (\Phi^j)' - 2 \pd_k G_{ij} (\Phi^k)' (\Phi^i)' (\Phi^j)' + 4 (\Phi^j)' \left[ V_j - G_{ij} (\Phi^i)'' \right] \,\, ,
\ee
where in the last term we have combined the third and fifth terms in (\ref{LdN3}) for convenience. Now, from $E1$ we have:
\be
V_j = G_{ij} (\Phi^i)'' + 4 G_{ij} A' (\Phi^i)' + G_{ij} {\cal G}^i{}_{kl} (\Phi^k)' (\Phi^l)' \,\, .
\ee
Substituting this into (\ref{LdN3p}), we are left with:
\be \label{LdN3pp}
L_{dN3} = - 2 \pd_k G_{ij} (\Phi^k)' (\Phi^i)' (\Phi^j)' + 4 G_{ij} {\cal G}^i{}_{kl} (\Phi^j)' (\Phi^k)' (\Phi^l)' \,\, .
\ee
To rewrite the last term in a convenient form, note that the standard expression for the Christoffel symbols
\be
{\cal G}^i{}_{kl} = \frac{1}{2} G^{i j} \left( \pd_k G_{j l} + \pd_l G_{k j} - \pd_j G_{kl} \right) \,\, ,
\ee 
immediately implies:
\be
G_{ij} {\cal G}^i{}_{kl} = \frac{1}{2} \left( \pd_k G_{j l} + \pd_l G_{k j} - \pd_j G_{kl} \right) \,\, .
\ee
Using this, it is easy to see that the second term in (\ref{LdN3pp}) becomes:
\be
4 G_{ij} {\cal G}^i{}_{kl} (\Phi^j)' (\Phi^k)' (\Phi^l)' = 2 \pd_k G_{ij} (\Phi^i)' (\Phi^k)' (\Phi^j)' \,\, .
\ee
Hence, we finally find that
\be \label{dN3zero}
L_{dN3} = 0
\ee
identically. Note that to reach this result we relied on the assumption that $A \neq const$, otherwise $A''$ would not appear in (\ref{LdN3}). So our derivation of (\ref{dN3zero}) is valid only for nontrivial warp factors $A(z)$. Clearly, if $A=const$ the system (\ref{E123}) simplifies right away. In fact, since $G_{ij}$ is diagonal, it is obvious that for $A =const$ and $\Lambda > 0$ there is no solution as equations $E2$ and $E3$ are incompatible.  

The result (\ref{dN3zero}) shows that, for nontrivial warp factor $A$, equations $E1$, $N2$ and $dN3$ (i.e. the derivative of $N3$) are not independent. We chose, for convenience, to use $E1$ and $N2$ inside $dN3$, in order to obtain that $dN3$ is identically satisfied. However, we could have equally well solved algebraically for $A''$ from $dN3$, i.e. (\ref{LdN3}) $\!=\!0$, and substituted the result in $N2$. The subsequent manipulations would have been the same. In other words, we can interpret the above dependency between the field equations as the statement that equation $N2$ is solved automatically, whenever equations $E1$ and $N3$ are (clearly, if $N3$ is satisfied, then so is $dN3$). Hence $N2$ can be dropped, leaving us with the system
\bea \label{EoM2}
(\Phi^i)'' + 4 A' (\Phi^i)' + {\cal G}^i{}_{jk} (\Phi^j)' (\Phi^k)' - G^{ij} V_j &=& 0 \nn \\
-4 \Lambda e^{-2A} + 12 A'^2 + 4 V - 2 G_{ij} (\Phi^i)' (\Phi^j)' &=& 0 \,\, .
\eea 
Therefore, there are equal numbers of equations and of unknown functions. This is a rather important statement as the metric ansatz (\ref{5dMetr})-(\ref{dSm})/(\ref{AdSm}), which we want for physical reasons, at first sight seemed to have one function less than needed. So it seemed that it might be (near) impossible to find solutions with that ansatz. Now, however, we see that the number of functions is just right.

\subsection{Simplification in scalar sector}

The system (\ref{EoM2}) is still rather daunting to address in full generality. However, in this subsection we will show that it is consistent to set three of the six scalar fields $\Phi^i (z)$ to zero. The resulting simplification will be of crucial importance for finding analytic solutions in the following. 

To understand which scalars one can set to zero in the full coupled system, let us first write down more explicitly all scalar field equations, namely: 
\be \label{EoMPhii}
\nabla^2 \Phi^i + {\cal G}^i{}_{jk} (\Phi^j)' (\Phi^k)' - G^{ij} V_j = 0 \,\, . 
\ee
Using (\ref{SigmaMCh}), equations (\ref{EoMPhii}) acquire the form:
\bea \label{6EoMs}
&&\nabla^2 p - V^p = 0 \,\,\, , \nn \\
&&\nabla^2 x + {\cal G}^x_{bb} (b')^2 - V^x = 0 \,\,\, , \nn \\
&&\nabla^2 g + {\cal G}^g_{aa} (a')^2 - V^g = 0 \,\,\, , \nn \\
&&\nabla^2 \phi + {\cal G}^{\phi}_{bb} (b')^2 - V^{\phi} = 0 \,\,\, , \nn \\
&&\nabla^2 a + {\cal G}^a_{ga} \,g' a' - V^a = 0 \,\,\, , \nn \\
&&\nabla^2 b + {\cal G}^b_{\phi b} \,\phi' b' + {\cal G}^b_{xb} \,x' b' - V^b = 0 \,\,\, ,
\eea
where $\nabla^2$ is as in (\ref{Nab2}). Now, from (\ref{Pot}) one can see that all terms in $V^a$ and $V^b$ are proportional to at least one power of $a$ or of $b$. Thus, if we take $a\equiv 0$ and $b\equiv 0$, then both the $\nabla^2 a$ equation and the $\nabla^2 b$ equation in (\ref{6EoMs}) will be identically satisfied. Similarly, one can find from (\ref{Pot}) that $V^{g} = 0$, when $g=0$ and $a= 0$. Therefore, the $\nabla^2 g$ equation of motion in (\ref{6EoMs}) is identically satisfied when $g$ and $a$ vanish. Summarizing, we can consistently set:
\be \label{gabzero}
g = 0 \quad , \quad a = 0 \quad , \quad b = 0
\ee
and drop the $\nabla^2 g$, $\nabla^2 a$ and $\nabla^2 b$ equations of motion.\footnote{This is a convenient place to comment on taking $H_3 = 0$ in the ansatz of Section 2. That this is consistent seems well known in the literature. However, we have not seen a more detailed discussion, like the one presented here for (\ref{gabzero}). So let us briefly outline why this is the case. If $H_3 \neq 0$, then according to (3.8) of \cite{BHM} there are two more independent scalars to consider, $h_1$ and $h_2$ in their notation. So now $\Phi^i=\{p,x,g,\phi,a,b,h_1,h_2\}$. The components of the sigma model metric $G_{ij} (\Phi)$ do not depend on $h_{1,2}$ and there are no nonzero mixed components of the form $G_{h_{1,2} \,\Phi^{\hat{i}}}$ with $\Phi^{\hat{i}} \neq h_{1,2}$; see (3.13) of \cite{BHM}. Hence all additional nonvanishing components of the Christoffel symbols are of the form ${\cal G}^{\Phi^i}_{\Phi^j h_{1,2}}$. Finally, $V^{h_{1,2}}|_{h_1=0=h_2} = 0$, as can be seen from (3.14) of \cite{BHM}. So setting $h_{1,2} = 0$ satisfies the $\nabla^2 h_{1,2}$ field equations and does not change at all the considerations regarding (\ref{6EoMs}).} Then the system (\ref{6EoMs}) reduces to:
\bea \label{EoMspxphi}
&&p'' + 4 A' p' - V^p = 0 \,\,\, , \nn \\
&&x'' + 4 A' x' - V^x = 0 \,\,\, , \nn \\
&&\phi'' + 4 A' \phi' - V^{\phi} = 0 \,\,\, .
\eea
Note also that, upon imposing (\ref{gabzero}), the last field equation in (\ref{EoM2}) becomes:
\be \label{EoMm}
4 \Lambda e^{- 2A} - 12 A'^2 + 12 p'^2 + 2 x'^2 + \frac{1}{2} \phi'^2 - 4 V = 0 \,\, .
\ee
Finally, substituting (\ref{gabzero}) into the potential (\ref{Pot}), we find:
\be \label{Vs}
V = \frac{e^{-4p-4x}}{4} - e^{2p-2x} + \frac{e^{8p-2x+\phi}}{4} P^2 + \frac{e^{8p-4x}}{8} Q^2 \,\, .
\ee

\section{Solutions with $\mathbf{dS_4}$ spacetime}
\setcounter{equation}{0}

In the following, we will adopt (\ref{gabzero}). Thus our goal will be to solve the system (\ref{EoMspxphi})-(\ref{EoMm}) with the potential as in (\ref{Vs}). To gain better understanding, we would like to find analytic solutions. For that purpose, it is rather useful to work in an approximation, in which one of the terms in (\ref{Vs}) dominates over the others. 

One limit that immediately comes to mind is taking $P>\!\!>1$, since in microscopic realizations of the duality $P = \frac{N_c}{4}$ (see, for instance, \cite{NPP}) with $N_c$ being the number of D5 branes sourcing the background. Alternatively, one could take $Q>\!\!>1$, where $Q$ is the $F_5$ flux due to D3 brane sources. However, there is a wider range of possibilities that may contain more interesting examples. To explain that, let us redefine the scalars $p,x,\phi$ as:
\be
p (z) = \tilde{p} (z) + p_0 \quad , \quad x (z) = \tilde{x} (z) + x_0 \quad , \quad \phi (z) = \tilde{\phi} (z) + \phi_0 \,\, ,
\ee
where $p_0 = const$, $x_0 = const$ and $\phi_0 = const$. Then (\ref{Vs}) becomes:
\be \label{VN}
V = \frac{1}{N_p^4 N_x^4} \,\frac{e^{-4\tilde{p}-4\tilde{x}}}{4} - \frac{N_p^2}{N_x^2} \,e^{2\tilde{p}-2\tilde{x}} + \frac{N_p^8 N_{\phi} P^2}{N_x^2} \,\frac{e^{8\tilde{p}-2\tilde{x}+\tilde{\phi}}}{4} + \frac{N_p^8 Q^2}{N_x^4} \,\frac{e^{8\tilde{p}-4\tilde{x}}}{8} \,\, .
\ee
where for convenience we have denoted
\be
N_p \equiv e^{p_0} \quad , \quad N_x \equiv e^{x_0} \quad, \quad N_{\phi} \equiv e^{\phi_0} \,\, .
\ee
Clearly then, we can pick different terms in (\ref{VN}) to be dominant by taking suitable limits for different ratios of the constants $N_p$, $N_x$, $N_{\phi}$, $P$ and $Q$. This is rather similar to the walking solutions of \cite{NPP}. Namely, there is an integration constant, denoted by $c$ there, such that the walking backgrounds are approximate solutions in the limit $\frac{c}{N_c} >\!\!> 1$ .

\subsection{Explicit solution} \label{ExplSol}

It turns out that a de Sitter solution (i.e. with $\Lambda > 0$), which is our primary interest, can be found analytically in the limit in which the second term in (\ref{VN}) is dominant. Clearly, one condition to achieve such a limit is
\be
\frac{N_p}{N_x} >\!\!> \frac{1}{N_p^2 N_x^2} \,\, ,
\ee
which implies the constraint
\be
\frac{1}{N_x} <\!\!< N_p^3 \,\, .
\ee
Another condition, from comparing the second and third terms in (\ref{VN}), is
\be
\frac{N_p}{N_x} >\!\!> \frac{N_p^4 N_{\phi}^{1/2} P}{N_x}  \qquad \Rightarrow \qquad N_p^3 <\!\!< \frac{1}{N_{\phi}^{1/2} P} \,\,\, .
\ee
Finally, comparing the second and fourth terms, we have:
\be
\frac{N_p}{N_x} >\!\!> \frac{N_p^4 Q}{N_x^2} \qquad \Rightarrow \qquad N_p^3 <\!\!< \frac{N_x}{Q} \,\,\, .
\ee
To summarize, by assuming
\be \label{Limits}
\frac{1}{N_x} <\!\!< N_p^3 <\!\!< \, {\rm smaller\,\,of} \, \left\{ \frac{1}{N_{\phi}^{1/2} P} \,\, , \,\, \frac{N_x}{Q} \right\} \,\, ,
\ee
we ensure that the potential (\ref{VN}) simplifies to:
\be \label{Vapprox}
V \approx - \,\frac{N_p^2}{N_x^2} \,e^{2\tilde{p}-2\tilde{x}} \,\, .
\ee
Let us for convenience denote $N^2 \equiv N_p^2 / N_x^2$ and drop the tildes from now on. So we write (\ref{Vapprox}) as:
\be \label{Vapp}
V \approx - \,N^2 \,e^{2p-2x} \,\, .
\ee

Note that one could take either $P=0$ or $Q=0$ in (\ref{VN}), in which case the $\{ , \}$ bracket in (\ref{Limits}) simplifies in the obvious manner. Also, since we are considering $H_3 = 0$ identically, we should make the following remark. To have a solution with nonvanishing $F_5$ flux, i.e. with $Q\neq 0$, in our set-up, one would need to have vanishing $F_3$ flux (which is guaranteed by taking $P=0$), in order to satisfy the 10d equation of motion $d \star (e^{-\Phi} H_3) = - F_5 \wedge F_3$. Such a solution, supported by constant $F_5$ flux, would be similar to the background used in \cite{GIN}. The solutions we will find below, however, all have $Q=0$. So they belong to the same class of solutions as \cite{MN} and its later deformations \cite{NPP,ENP}. 

\subsubsection{Reducing the scalar sector} \label{ScS}

From (\ref{Vapp}) we find that:
\be \label{Vder}
V_p \approx - 2 e^{2p - 2x} N^2 \quad , \quad V_x \approx 2 e^{2p - 2x} N^2 \quad , \quad V_{\phi} \approx 0 \,\, .
\ee
The vanishing of $V_{\phi}$ immediately implies that we can solve the third equation in (\ref{EoMspxphi}) by taking
\be \label{phis}
\phi = 0 \,\, .
\ee
In addition, from (\ref{Vder}) we see that:
\be
V_x = - V_p \,\, ,
\ee
which upon using (\ref{SigmaMM}) gives:
\be
V^x = - 6 V^p \,\, . 
\ee
It is clear then that, by taking
\be \label{xp}
x = - 6 p \,\, ,
\ee
we can ensure that the field equation $x'' + 4 A' x' - V^x = 0$ becomes exactly the same as $p'' + 4 A' p' - V^p = 0$. So, adopting (\ref{xp}), we reduce the system to a single scalar $p(z)$ with the potential $V = - N^2 e^{14 p}$.

\subsubsection{Solving the field equations} \label{EoMsol}

Substituting (\ref{phis}) and (\ref{xp}) into (\ref{EoMspxphi})-(\ref{EoMm}), we obtain the system:
\bea \label{syst}
p'' + 4 A' p' + \frac{N^2}{3} e^{14p} &=& 0 \nn \\
4 \Lambda e^{- 2A} - 12 A'^2 + 84 p'^2 + 4 N^2 e^{14p} &=& 0 \,\, .
\eea
We will leave for the future solving the above system (analytically) in full generality and, instead, will find here a particular solution of a rather simple form. To do that note that there are only two kinds of exponentials, $e^{-2A}$ and $e^{14p}$, in this system. So it seems logical to expect that it may be possible to find a solution of (\ref{syst}), when those two exponentials are equal to each other up to a constant. Hence, let us make the ansatz:
\be \label{Apc0}
A(z) = - 7 p(z) + c_0 \,\, ,
\ee
where $c_0 = const$. Substituting (\ref{Apc0}) into (\ref{syst}), we find:
\bea \label{syst2}
p'' - 28 p'^2 + \frac{N^2}{3} e^{14p} &=& 0 \nn \\
\left( \frac{4 \Lambda}{e^{2 c_0}} + 4 N^2 \right) e^{14p} - 504 \,p'^2 &=& 0 \,\, .
\eea
Let us for convenience introduce the notation
\be \label{alphabetaP}
\alpha = \frac{4 \Lambda}{e^{2 c_0}} + 4 N^2 \qquad {\rm and} \qquad \beta = 504 \,\, ,
\ee
so that the second equation in (\ref{syst2}) can be written as:
\be \label{Eab}
\alpha \,e^{14p} - \beta \,p'^2 = 0 \,\, .
\ee
Solving this algebraically for $e^{14 p}$ in terms of $p'^2$ and substituting the result in the first equation of (\ref{syst2}), we find:
\be \label{Eg}
p'' - \gamma \,p'^2 = 0 \,\, ,
\ee
where we have denoted 
\be \label{gammaC}
\gamma = 28 - \frac{N^2}{3} \frac{\beta}{\alpha} \,\, .
\ee
Now the system we need to solve is (\ref{Eab})-(\ref{Eg}). The question is whether we can choose the integration constants so that both equations are satisfied simultaneously. 

The general solution of (\ref{Eab}) is:
\be \label{p1}
p_1(z) = \frac{1}{14} \, \ln \!\left[ \frac{\beta}{49 \,\alpha \,(z - C_1)^2} \right] \, ,
\ee
where $C_1$ is an integration constant. For later convenience, let us write this as:
\be \label{p1p}
p_1 (z) = - \frac{1}{14} \left[ \ln \left( \frac{49 \,\alpha}{\beta} \right) + 2 \ln (z - C_1) \right] \,\,.
\ee
The general solution of (\ref{Eg}) is:
\be
p_2(z) = - \frac{1}{\gamma} \,\ln \!\left( \gamma z C_2 + \gamma C_3 \right) \,\, ,
\ee
where $C_2$ and $C_3$ are integration constants. Since the latter are arbitrary, we can rescale them by another constant $\kappa$ and write:
\be \label{p2}
p_2 (z) = - \frac{1}{\gamma} \left[ \,\ln (\gamma \kappa) + \ln (z C_2 + C_3) \,\right] \,\, .
\ee
Let us now compare (\ref{p2}) to (\ref{p1p}). Clearly, the two expressions can be made equal if we take:
\be \label{C3kappaP}
C_2 = 1 \quad , \quad C_3 = - C_1 \equiv C \quad , \quad \gamma = 7 \quad , \quad \kappa = \left( \frac{\alpha}{\beta} \right)^{\!1/2} \,\, .
\ee
However, unlike $C_{1,2,3}$ and $\kappa$, which until (\ref{C3kappaP}) were arbitrary, $\gamma$ was already given in (\ref{gammaC}). So we need to satisfy one more condition, namely:
\be \label{7alphabetaP}
7 = 28 - \frac{N^2}{3} \frac{\beta}{\alpha} \,\,\, ,
\ee
in order to have a solution. Substituting the definitions of $\alpha$ and $\beta$ from (\ref{alphabetaP}), the relation (\ref{7alphabetaP}) becomes: 
\be \label{Hc0}
\Lambda = e^{2c_0} N^2 \,\, .
\ee
Clearly, this condition means that $\Lambda$ is positive-definite and thus the 4d spacetime is de Sitter.

\subsubsection{Summary and discussion}

To summarize, we have found the solution:
\bea \label{PsolpA}
p(z) &=& - \frac{1}{7} \,\ln (z+C) - \,\frac{1}{14} \,\ln \!\left( \frac{7 N^2}{9} \right) \,\, , \nn \\
A(z) &=& \ln (z+C) \,+ \,\frac{1}{2} \,\ln \!\left(\frac{7 \Lambda}{9}\right) \,\, ,
\eea
where in the constant term of $A$ we have combined the constant coming from $p(z)$ with the contribution of $c_0$ as determined by (\ref{Hc0}). One can easily verify that this solution is indeed correct by substituting (\ref{PsolpA}) into (\ref{syst}), or equivalently into (\ref{E123}) with $g=0$, $a=0$, $b=0$, $\phi=0$, $x=-6p$ and $V = -N^2 e^{14p}$.

Recall that in this set-up $z$ is a radial variable with a natural range $[0,\infty)$. Clearly, by choosing the integration constant $C>0$, we can ensure that the solution is regular at the origin $z=0$. Obviously, though, the expressions in (\ref{PsolpA}) diverge as $z$ tends to $\infty$. However, in physical applications it may be sensible to introduce an upper cut-off $z_c$, such that $z \in [0,z_c]$. This could be a (dynamical) physical scale above which the model, based on the above solution, is not valid. For example, if we view the $dS_4$ background corresponding to the  solution (\ref{PsolpA}) as an approximation to the Inflationary epoch in the Early Universe, then a natural choice for $z_c$ is the scale of onset of Inflation, or perhaps the Hubble scale. Of course, when studying fluctuations in the above background, perturbative stability will depend on the boundary conditions one imposes at $z_c$. We leave this very interesting issue for a future investigation.

Let us also comment on whether (\ref{PsolpA}) is valid in the approximations within which we derived it, as should be the case. Such a question may arise at first sight, because the solution for $p(z)$, and thus also for $x(z)$, depends on $N$, which could be large. Since the dominance of the second term in (\ref{VN}) was obtained by imposing constraints on the coefficients $N_{p,x,\phi}$ only, it is natural to ask whether factors of $N$ coming from the exponentials of $p(z)$ and $x(z)$ could spoil this approximation. Fortunately, one can easily verify that this does not happen. Indeed, in the first term in (\ref{VN}) we have\footnote{As before, we drop the tildes.}: $e^{-4p-4x}|_{x=-6p} = e^{20 p} \sim \frac{1}{N^{20/7}} = \frac{1}{N^{2.857}}$; in the second term: $e^{2p-2x}|_{x=-6p} = e^{14p} \sim \frac{1}{N^2}$; in the third one: $e^{8p-2x}|_{x=-6p} = e^{20p} \sim \frac{1}{N^{2.857}}$; and, finally, in the fourth term: $e^{8p-4x}|_{x=-6p} = e^{32p} \sim \frac{1}{N^{4.57}}$. So even among the exponentials alone, for large $N$, the exponential of the second term dominates when we have (\ref{PsolpA}) and (\ref{xp}). Hence this is a valid solution. 

Finally, let us note that this solution shares important features with the ${\cal N} = 1$ supersymmetric walking backgrounds of \cite{NPP}. The letter are deformations of the Maldacena-Nunez solution \cite{MN}. As such, they naturally belong to the class of asymptotically linear dilaton (ALD) backgrounds. Recall that the linear dilaton behavior $\phi \sim z$ refers actually to a different coordinate system (in string frame) than the conventional one (in Einstein frame), used in \cite{MM} to establish the holographic renormalization of ALD backgrounds. For comparison of the two systems, see for example \cite{MV}. In the coordinate system of \cite{MM}, that is relevant for us, the asymptotic behavior of the dilaton is $\phi \sim c_1 \ln z + c_2 + ...$\,, where $c_{1,2} = const$. Now, for the walking solutions of \cite{NPP} the dilaton is constant, while it is zero here. Hence, both these walking backgrounds, as well as our de Sitter foliation one, exhibit special cases of ALD behavior with a dilaton solution having $c_1 = 0$. Of course, in addition to the dilaton, we have another nontrivial scalar, which is diverging in a manner consistent with the ALD asymptotics, namely $p(z) \sim \ln z$. This is the case for all of the backgrounds of \cite{NPP}-\cite{EGNP} (see also \cite{HNP}) too, albeit with different divergent scalars in each case. More broadly, one should keep in mind that, while $AdS$ asymptotics means that the gravity background is dual to a fundamental field theory, divergences (as in the ALD case, or more severe) signify that the gravity dual describes an effective field theory, which needs a UV completion. For a nice discussion on that point, one can see for example \cite{CGNPR}.

The backgrounds of \cite{NPP} also belong to the class of solutions found within the consistent truncation of \cite{BHM}, as already mentioned in Section 2. More importantly, in the walking region of \cite{NPP}, the scalars $a$, $b$ and $g$ vanish (see \cite{ASW}), just as they do in the new solution here. Also, the string dilaton, $\phi$, is constant in both cases, as mentioned above. And, of course, our solution, when lifted to a full ten-dimensional one, has exactly the same topology of the compact 5d space, namely $S^1 \times S^2 \times S^2$, as in \cite{NPP}, due simply to the structure of the metric ansatz (\ref{10dmetric}). Therefore, it seems tempting to speculate that the $dS_4$ foliation found here could be obtained by a certain nonsupersymmetric deformation of the ${\cal N}=1$ walking backgrounds of \cite{NPP}. Understanding this issue would undoubtedly be very important for finding microscopic (D-brane) realizations of our solution. We leave this investigation for the future.  

\subsection{Solutions for large $P$}

The kind of reduction of the scalar sector to a single independent field, that we found in Subsection \ref{ScS}, is not unique to the limit in which the second term in (\ref{VN}) dominates. In fact, a similar reduction can be obtained for any of the four terms of the potential being the leading one. Despite that, we did not find three more analogues of the solution (\ref{PsolpA}). It is instructive to understand why. So, following Section \ref{ExplSol}, we will outline the relevant considerations for the case of another leading term in (\ref{VN}).

Let us take the approximation
\be \label{VP}
V \approx \frac{e^{8p-2x+\phi}}{4} P^2 \,\, ,
\ee
where as before we have dropped the tildes. Clearly, the simplest, although certainly not the only, way to ensure (\ref{VP}) is to take 
\be
P >\!\!> 1 \qquad , \qquad Q = 0
\ee
and $N_{p,x,\phi} \sim {\cal O} (1)$ in (\ref{VN}). Note that any overall $N$ factor in (\ref{VP}) just rescales the parameter $P$. So, for our purposes, there is no loss of generality by not including such a factor in the potential here.

As in Section \ref{ExplSol}, from (\ref{VP}) we can deduce that setting
\be
x = - \frac{1}{2} \phi
\ee
makes the equations of motion of $x$ and $\phi$ equivalent. As a result, we are left with the following scalar field equations:
\be
\nabla^2 \phi = e^{8p+2\phi} P^2 \qquad , \qquad \nabla^2 p = \frac{1}{3} \,e^{8p+2\phi} P^2 \,\, .
\ee
Adding and subtracting these two equations, we find
\be
\nabla^2 \left( 8p + 2 \phi \right) = \frac{14}{3} e^{8p+2\phi} P^2 \qquad {\rm and} \qquad \nabla^2 \left( \phi - 3p \right) = 0 \,\,\, .
\ee
Obviously, the second equation above can be solved by taking
\be \label{ph3p}
\phi = 3p \,\, .
\ee 
Hence, the system (\ref{EoMspxphi})-(\ref{EoMm}) now becomes:
\bea \label{systPn}
p'' + 4 A' p' - \frac{1}{3} \,e^{14p} P^2 &=& 0 \nn \\
4 \Lambda e^{-2 A} - 12 A'^2 + 21 p'^2 - e^{14p} P^2 &=& 0 \,\, .
\eea

At this stage, one could make the ansatz
\be
A(z) = - 7 p(z) + const \,\, ,
\ee
as in (\ref{Apc0}). However then, going through the same steps as in Subsection \ref{EoMsol}, one finds that the analogue of the last condition, i.e. equation (\ref{7alphabetaP}), cannot be satisfied. The same problem occurs for the limits, in which any of the remaining two terms in (\ref{VN}), i.e. the first or the last, is the dominant one.

To understand what is the key difference that allowed the existence of the solution (\ref{PsolpA}), let us compare (\ref{syst}) and (\ref{systPn}). We see that both systems are of the form:
\bea \label{systeta}
p'' + 4 A' p' - \eta \,\omega\,L^2 e^{\,\zeta \,p} &=& 0 \nn \\
4 \Lambda e^{-2A} - 12 A'^2 + \xi p'^2 - 3 \,\eta\,L^2 e^{\,\zeta \,p} &=& 0 \,\, ,
\eea
where $\eta$, $\zeta$, $\xi$ and $\omega$ are numerical coefficients and the parameter $L$ denotes $N$ or $P$. In fact, the coefficient $\xi$ has to satisfy $\xi = \frac{3 \zeta}{2 \omega}$ due to its origin from the system (\ref{E123}). Now, substituting the ansatz $A(z) = - \frac{\zeta}{2} \,p(z) + c_0$ in (\ref{systeta}) and performing the same steps as in Subsection \ref{EoMsol}, we find again the system $\alpha e^{\,\zeta \,p} - \beta p'^2 = 0$ and $p'' - \gamma p'^2 = 0$, but with different coefficients $\alpha$, $\beta$ and $\gamma$; in particular: 
\be
\alpha \equiv \frac{4\Lambda}{e^{2c_0}} - 3 \,\eta L^2 \,\, . 
\ee
Finally, the analogue of (\ref{Hc0}) gives:
\be \label{LamGen}
\Lambda = \frac{(2-\zeta \omega) \,\eta \,e^{2 c_0} L^2}{2} \,\, .
\ee
When the potential is approximated by any of the other terms, except for the second one in (\ref{VN}), it turns out that we always have $2 < \zeta \omega$ with $\eta > 0$, and also $\beta > 0$. This would seem to suggest that there is a solution with $\Lambda < 0$, i.e. $AdS_4$ spacetime. However, (\ref{LamGen}) was derived under the assumption that $\alpha e^{\,\zeta \,p} - \beta p'^2 = 0$ has a solution, which for $\beta > 0$ is only possible if $\alpha > 0$. The latter inequality, though, cannot be satisfied for $\Lambda < 0$, since $\eta > 0$. The crucial difference  for the second term in (\ref{VN}), compared to the above considerations, is its overall minus sign. This leads to $\eta < 0$, while still $2 < \zeta \omega$. So now one can have $\Lambda > 0$ from (\ref{LamGen}), while $\alpha > 0$. This gives precisely the solution (\ref{PsolpA}).

Although the system (\ref{systPn}) does not have a special solution similar to (\ref{PsolpA}), we can still extract certain information analytically. This will enable us to find a class of  interesting numerical solutions to (\ref{systPn}). We turn to these considerations next. 

\subsubsection{Reduction to single equation}

Since our main interest is in de Sitter space, we would like to look for solutions with $\Lambda > 0$. So let us for convenience introduce the Hubble parameter:
\be 
H^2 \equiv \frac{\Lambda}{3} \,\,\, ,
\ee
such that the scale factor in (\ref{dSm}) becomes the standard $e^{Ht}$. 

Now, we can rewrite the system (\ref{systPn}) in a more convenient form by introducing $B(z)$ and $q(z)$ via:
\be \label{sub1}
A(z) = \ln \left( \frac{H}{B(z)} \right) \qquad {\rm and} \qquad p(z) = \frac{1}{7} \ln\left( \frac{q(z)}{P} \right) \,\, .
\ee
Then (\ref{systPn}) becomes:
\bea\label{SystF3}
\frac{q''}{q^2\,q'}&=&\frac{q'}{q^3}+\frac{7\,q}{3\,q'}+\frac{4\,B'}{q^2 B} \,\, ,\nn\\
\frac{3\,q'{}^2}{7\,q^2}&=&\frac{12\,B'{}^2}{B^2}+q^2-12\,B^2 \,\, .
\eea
Clearly, from the first equation above we can solve algebraically for $\frac{\,B'}{B}$ and substitute the result in the second equation. This allows us to solve for $B$:
\be \label{Bsol}
B = \frac{1}{2\,\sqrt{3}}\,\sqrt{\frac{(7\,q^4+3\,q'{}^2-3\,q\,q'')^2}{12\,q^2\,q'{}^2}+q^2-\frac{3\,q'{}^2}{7\,q^2}} \,\, .
\ee 
Now, substituting (\ref{Bsol}) into the first equation of (\ref{SystF3}), we obtain a third order factorized differential equation for $q$ of the form $F_1\times F_2=0$, where
\bea\label{beta}
F_1&:& \,\, q''=\frac{q'{}^2}{q}+\frac{7}{3}\,q^3,\nn\\
F_2&:& \,\, q^{\prime \prime \prime}=\frac{49\,q^6}{36\,q'}+\frac{17}{2}\,q^2\,q'-\frac{9\,q'{}^3}{28\,q^2}-\frac{7\,q^3\,q''}{2\,q'}+\frac{q'\,q''}{2\,q}+\frac{5\,q''{}^2}{4\,q'} \,\, .
\eea
So we have reduced solving the coupled system of differential equations (\ref{systPn}) to solving either of the ODEs $F_1$ and $F_2$. 

Actually, $F_1$ can be solved easily, giving:
\be \label{qsol}
q(z) = \sqrt{\frac{3}{7}} \,\frac{C_1}{\sin \left[ (z + C_2) C_1 \right]}
\ee
with $C_{1,2}$ being integration constants. Now, substituting (\ref{qsol}) into (\ref{Bsol}), we find:
\be \label{Bsol2}
B = \frac{C_1}{2\sqrt{7}} \,\, .
\ee
Hence, from (\ref{sub1}) it follows that the warp factor $A$ is constant. However, recall from Section \ref{DepEq} that there is no solution for $\Lambda > 0$ and $A=const$. Indeed, one can verify that the third equation (i.e. $N2$ in (\ref{N2N3}), which for $A=const$ is independent of the other two) is not satisfied for the putative solution given by (\ref{qsol}) and (\ref{Bsol2}). Therefore, solving the system (\ref{systPn}) reduces to solving the single equation $F_2$ in (\ref{beta}).

\subsubsection{Investigating the solutions}

Since the differential equation $F_2$ does not depend explicitly on $z$ and is an odd function of $q$ and its derivatives, we can reduce its order by one via introducing the new independent variable $y=q(z)$ and the new function $R(y)=q'(z)^2$. Then $F_2$ becomes:\footnote{To verify this result, keep in mind that $\frac{dy}{dz} = q'$ and that $R' = \frac{dR}{dy}$.} 
\be \label{Req}
R'' = 17\,y^2 + \frac{49\,y^6}{18\,R} - \frac{9\,R}{14\,y^2} + \frac{R'}{2\,y} - \frac{7\,y^3 R'}{2\,R} + \frac{5\,R'{}^2}{8\,R} \,\, .
\ee
One can further simplify this equation by making the ansatz
\be\label{Rdef}
R = \frac{49}{9}\,y^4 \left[1+T(\ln y)\right] \,\, ,
\ee
where $T$ is an yet undetermined function. Substituting (\ref{Rdef}) into (\ref{Req}), we find the following equation for $T(w)$:
\be\label{T}
(1+T)\,T'' = \frac{5}{8}\,T'{}^2-\frac{36}{49}\,T-\frac{9}{14}\,T^2-\frac{15}{7}\,T'-\frac{3}{2}\,T\,T' \,\, .
\ee

Note that this equation is invariant under a translation $w \rightarrow w - w_0$ with $w_0$ being a constant. Since, according to (\ref{Rdef}), one has $w = \log (q)$, the $w$ translation is equivalent to a $q$ rescaling of the form $q\to q\,/\,q_0$ with $q_0 = e^{w_0}$. In fact, it is useful to rewrite (\ref{Rdef}) as:
\be\label{betaeq}
q'(z)= \pm \frac{7}{3} \,q^2 \,\sqrt{1+T} \,\, ,
\ee 
where we have used that $y=q$ and $R=q'{}^2$. Now it is easy to realize that a rescaling $q\to q\,/\,q_0$ leaves (\ref{betaeq}) invariant, if it is accompanied by the transformation $z\to z\,q_0$. In other words, equations (\ref{T}) and (\ref{betaeq}) are invariant under the simultaneous rescalings:
\be \label{rescale}
q\to q\,/\,q_0 \equiv \hat{q} \qquad {\rm and} \qquad z\to z\,q_0 \equiv \hat{z} \,\, .
\ee
Hence, one can generate new solutions $\hat{q} (\hat{z})$ by performing these rescalings on a known solution $q(z)$. This will play an important role below.

\begin{figure}[htbp]
\begin{center}
\includegraphics[width=3.5in]{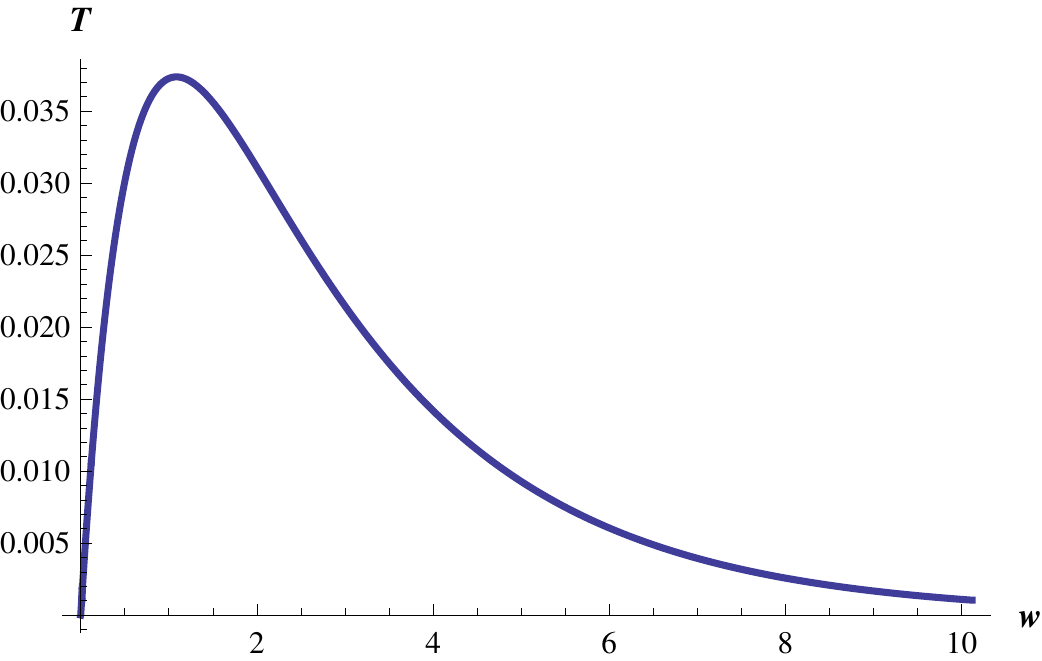}
\caption{The function $T(w)$ for integration constant $t_1 = 0.1$.}
\label{Twplot}
\end{center}
\end{figure}

Let us now focus on investigating equation (\ref{T}). Unfortunately, it cannot be solved exactly by analytical means. Nevertheless, we will be able to find numerical solutions. As a first step in doing that, we need the series expansion of $T$ around some point, which we choose for convenience to be $w=0$. To simplify the considerations, we will look for solutions, such that the function $T(w)$ has a zero somewhere.\footnote{Note that there are plenty of functions that do not have any zeros, like for example $\cosh (w)$.} This assumption implies that the Taylor expansion in our case has the form $T(w) = t_1 w + t_2 w^2 + ...$ with $t_{1,2} = const$. Indeed, the generic expansion would be $T(w) = t_0 + t_1 w + ...$\,, where $t_1 \neq 0$ due to the assumption of the presence of a zero of $T(w)$. Therefore, recalling that a translation of $w$ is a symmetry of equation (\ref{T}), we can use the shift $w\to w-\frac{t_0}{t_1}$ to set to zero the constant term. Once we find a solution with $t_0 = 0$, the more general solution, containing an additional integration constant $w_0$, can be obtained by performing the shift $w \to w - w_0$. So, substituting an expansion of the form $T(w) = t_1 w + ...$ in (\ref{T}), we find that the Taylor series of $T$ around $w=0$ is given by:
\be\label{Tser}
T(w)=t_1\,w\left[1+\frac{5}{112}(7\,t_1-24)\,w+\frac{1}{1344}(35\,t_1{}^2-756\,t_1+864)\,w^2+ ...\right] \, .
\ee
Using (\ref{Tser}) to set the initial conditions, we can solve (\ref{T}) numerically for any choice of the constant $t_1$. However, we show in the Appendix that the expression for $B(z)$, following from (\ref{Tser}), is real only for $t_1>0$ or $t_1 < - \frac{16}{7}$. Taking $t_1=0.1$\,, we plot in Figure \ref{Twplot} the function $T(w)$ obtained from the numerical integration.
Other choices of positive $t_1$ give very similar results. As evident from the figure, $T(w)$ increases until reaching the maximum value $T_{\rm max}\simeq 0.038$ at $w\approx1$ and then decreases to zero as $w\to\infty$.

\begin{figure}[htbp]
\begin{center}
\includegraphics[width=3.5in]{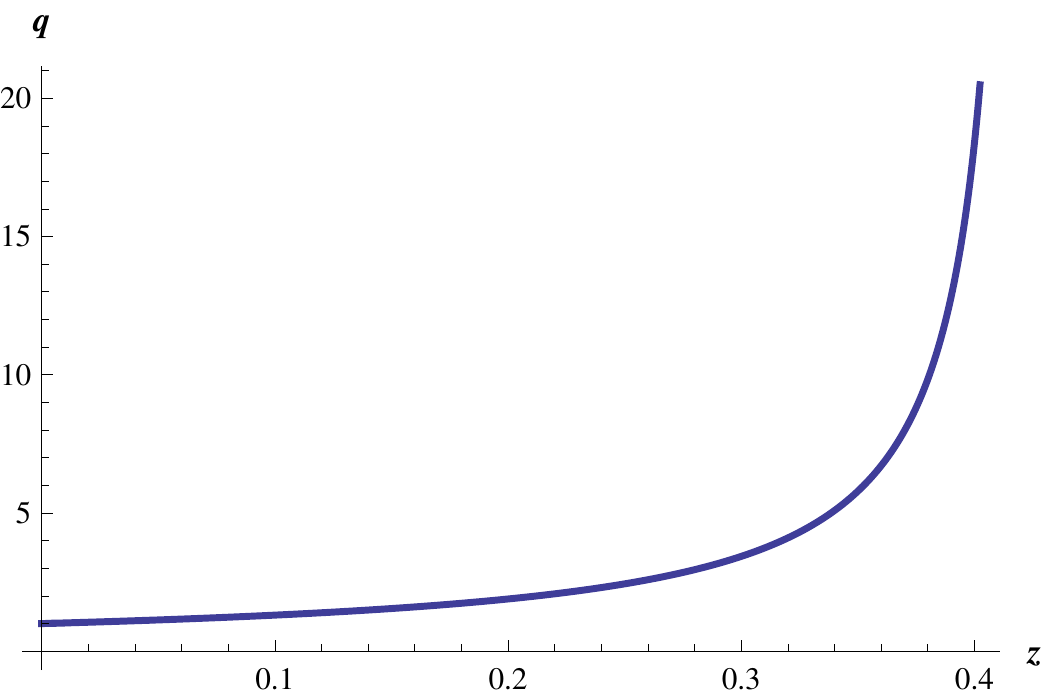}
\caption{The function $q(z)$ for integration constant $t_1 = 0.1$.}
\label{zqplot}
\end{center}
\end{figure}

Having a numerical solution for $T(w)$, we can now find numerically $q(z)$ by integrating (\ref{betaeq}). Indeed, recalling that $w=\log(q)$, we find from (\ref{betaeq}) {}\footnote{For convenience, here we take the plus sign in (\ref{betaeq}). It is easy to transform the results to the case with the minus sign. We will comment more on that at the end of this section.} :
\be\label{z}
z=\frac{3}{7}\int_0^w \frac{e^{-w}}{\sqrt{1+T(w)}}\,dw \,\, ,
\ee
where we have fixed the integration constant by taking $w=0$ at $z=0$. Note that requiring $z\ge 0$ implies that $w \ge 0$, or equivalently $q \ge 1$, in (\ref{z}). Substituting the numerical results for $T(w=\log q)$ into (\ref{z}), we can compute $z$ as a function of $q$. The result, plotted as the inverse function $q(z)$, is shown in Figure \ref{zqplot}. 
The function $q(z)$ diverges at $z_{\rm max}\simeq0.4231$. This upper bound on $z$ can be understood in the following way. In the limit $T\to 0$, the value of the integral for $w \to \infty$ tends to $3\,/\,7=0.4285$. Since $T$ is small, but nonzero, we have that $z_{\rm max}=0.4231< 3\,/\,7$. At first sight, the allowed range of $z$ seems very short. However, recall that the solution can be rescaled by the simultaneous transformations (\ref{rescale}). So, by taking $q_0>\!\!>1$, one can make $z_{\rm max}$, and thus the range of $z$, arbitrarily large. Note also, that another way of varying the range of $z$ is by changing the value of the constant $t_1$, as can be seen from (\ref{z}).

\begin{figure}[htbp]
\begin{center}
\includegraphics[width=3.5in]{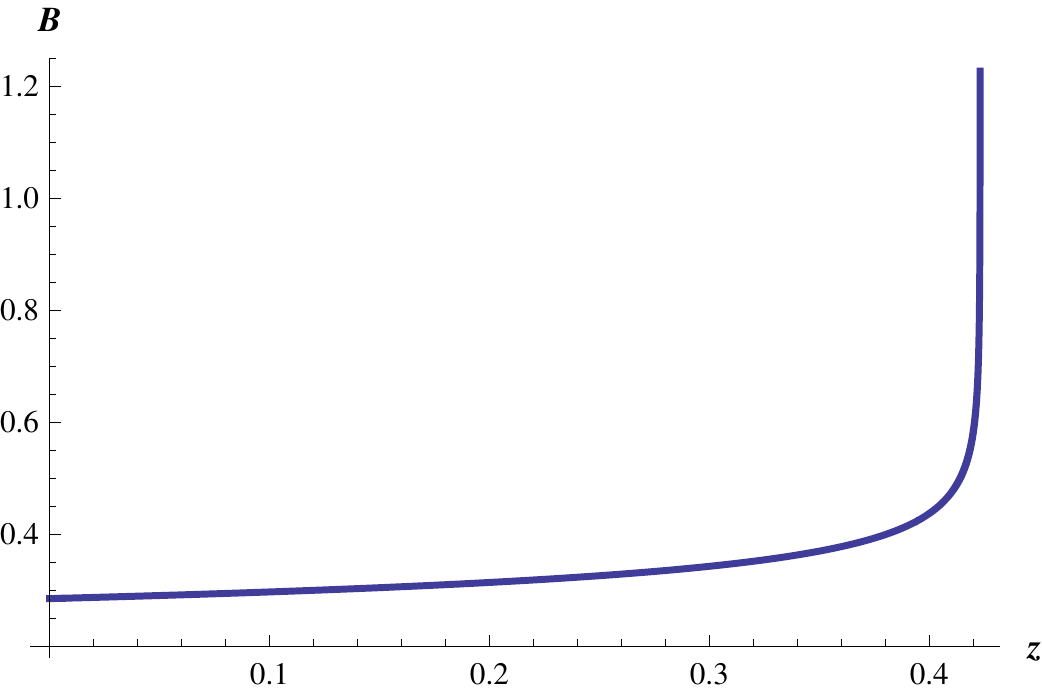}
\caption{The function $B(z)$ for integration constant $t_1 = 0.1$.}
\label{zBplot}
\end{center}
\end{figure}

Having found $q(z)$ numerically, we can compute numerically $B(z)$ as well by using (\ref{Bsol}). The function $B(z)$ is plotted in Figure \ref{zBplot}. 
It is finite at $z=0$ and diverges at $z_{\rm max}$. From (\ref{sub1}), we then conclude that the functions $p(z)$ and $A(z)$ are also regular at $z=0$ and have logarithmic singularities at $z_{\rm max}$.

Let us also comment on the behavior of the solution for $q(z)$, when the negative sign in (\ref{betaeq}) is taken. Let us write the result of the integration in this case as:
\be \label{z1}
z=-\frac{3}{7}\int_0^w \frac{e^{-w}}{\sqrt{1+T(w)}} \, dw =\frac{3}{7}\int_w^0 \frac{e^{-w}}{\sqrt{1+T(w)}} \, dw \,\, ,
\ee
where again the integration constant is chosen so that $q(0)=1$. Since now the positivity of $z$ implies negative $w$, we have that $ q(z) = e^w \le 1$. Choosing $t_1=0.1$ as before, we find, in an extension of our numerical computation to negative $w$, that in the limit $w\to w_{\rm min}= -1.757$ the function $T(w)\to -1$. Hence, for $w\to w_{\rm min}$ the integral in (\ref{z1}) diverges, thus allowing us to identify this limit with $z\to\infty$. Clearly then, the function $q(z)$ takes the finite value $q=e^{w_{\rm min}} \approx 0.17$ as $z\to\infty$. Also, obviously, we have $q=1$ at $z=0$. So, as $z$ runs over the whole real axis, $q(z)$ takes values in a finite range, namely: $e^{w_{\rm min}} \le q \le 1$. Therefore, from (\ref{sub1}) we can see that the scalar field $p(z)$ does not have a singularity in this case.

Finally, let us note that a discussion, similar to the one at the end of Section 4.1.3, applies here as well. More precisely, the numerical solutions found above have vanishing $g$, $a$ and $b$ scalars as before. Also, their lifts to 10d have exactly the same compact 5d topology as before for, obviously, the same reasons. However, now the string dilaton $\phi$ is not a constant. Indeed, recall from (\ref{ph3p}) that $\phi = 3 p$. Therefore, the second numerical solution, the one corresponding to the minus sign in (\ref{betaeq}), has finite $\phi$ for any $z$, in accord with the discussion below (\ref{z1}). It would be interesting to explore whether this solution can be obtained from a nonsupersymmetric deformation of the walking backgrounds of \cite{NPP}, or more generally \cite{ENP,EGNP}. On the other hand, the numerical solution, corresponding to the plus sign in (\ref{betaeq}), has a logarithmically divergent dilaton at the upper limit of the radial direction, as is clear from the discussion following (\ref{z}). Thus it seems unlikely that this solution could be related to deformations of walking backgrounds like \cite{NPP,ENP,EGNP}, which all have finite dilaton asymptotics. However, it is conceivable that it could have a connection to a more general (non-walking) solution of the full ${\cal N}=1$ system of \cite{CNP,HNP}. We hope to come back to those interesting issues in the future.

\section{Toward glueball inflation}
\setcounter{equation}{0}

Here we studied the equations of motion of the 5d action (\ref{S5d})-(\ref{Pot}), which is a consistent truncation of 10d type IIB, with the metric ansatz (\ref{5dMetr})-(\ref{4dR}) and with only $z$-dependent profiles for the scalars. We found solutions of the form 
\be \label{MetricdS}
ds_5^2 = e^{2 A(z)} \left( -dt^2 + e^{2 H t} d\vec{x}^2 \right) + dz^2 \,\, ,
\ee
where the Hubble parameter $H = \sqrt{\frac{\Lambda}{3}}$ and $\Lambda > 0$. In the context of the gauge/gravity duality, this gives a dual description of certain strongly coupled gauge theories in a 4d de Sitter spacetime, thus providing a powerful tool for studying the latter. This topic was already addressed in \cite{GIN}, although the set of nontrivial ten-dimensional fields in their case is not compatible with the consistent truncation of \cite{BHM}, that we have used here.

Studying field theory in $dS_4$ background is of great interest both because of Cosmological Inflation and also because the present day Universe has a small, but non-vanishing, positive cosmological constant. However, there are cosmological models with even more involved time dependence of the 4d spacetimes; see \cite{EGM}, for instance, for interesting examples of holographic duals of such models.\footnote{Note that the remark we made above, regarding the relation of \cite{GIN} to the consistent truncation of \cite{BHM}, applies to \cite{EGM} as well, as they use the same field space truncation as \cite{GIN}.} In fact, even for Inflation one would want a time-dependent Hubble parameter, albeit a very slowly varying one \cite{DB}. So, although (\ref{MetricdS}) can be viewed as a leading approximation in the context of Inflation, ultimately we would like to have a solution of the more general form
\be
ds_5^2 = e^{2 A(z)} \left[ -dt^2 + s(t)^2 d\vec{x}^2 \right] + dz^2
\ee
with the scale factor $s(t)$ satisfying $\ddot{s} > 0$, where we have denoted \,$\dot{} \,\equiv \frac{d}{dt}$. In this notation, the standard definition of the Hubble parameter is
\be
H \equiv \frac{\dot{s}}{s}
\ee
and the definition of slow variation (the ``slow roll" regime) is
\be
\varepsilon \equiv - \frac{\dot{H}}{H^2} <\!\!< 1 \,\, .
\ee
To find solutions of (\ref{EoM}) of this kind, one needs to modify the ansatz for the scalar fields in the following way:
\be
\Phi^i = \Phi^i (t,z) \,\, .
\ee
Then the field equations become much more complicated. Working with (\ref{gabzero}) provides some simplification. Nevertheless, it is quite a nontrivial task to find analytic solutions. So we leave the detailed investigation of this question for the future. 

It is worth, though, to expand here upon why such solutions would be of great value for Inflationary Cosmology. Recall that, when gravitational backgrounds solving (\ref{EoM}) (or, more precisely, the 10d type IIB action, which all solutions of this 5d system lift to) are viewed as dual descriptions of strongly coupled gauge theories, the scalars $\Phi^i$ represent the glueball bound states. So the fact, that time-dependence in (some of the) $\Phi^i$ drives the cosmological expansion of the 4d spacetime, means that the role of the 4d inflaton field is played here by a composite scalar, more precisely a glueball, in a strongly coupled gauge sector. Thus, this would be a gravitational dual of glueball inflation. In that regard, recall that knowing the Hubble parameter as a function of time, i.e. the function $H(t)$, is the only 4d spacetime dynamics one needs, in order to extract predictions for the key inflationary observables $n_s$ (scalar spectral index) and $r$ (tensor to scalar ratio), as the scalar and tensor power spectra are entirely determined by algebraic expressions in terms of $H(t)$ and $\dot{H}(t)$; see \cite{DB}. 

Models, in which the inflaton is composite instead of a fundamental scalar, have already been proposed within purely field-theoretic inflationary model building \cite{CJS,BCJS}.\footnote{In addition, a holographic model of composite inflation due to a quark condensate was proposed in \cite{EFK}, based on embedding flavor probe branes in $AdS_5 \times S^5$. The strongly coupled gauge theory in that case is ${\cal N}=4$ SYM with a small number of quark hypermultiplets.} The main motivation is that in such models the so called $\eta$-problem does not occur. This problem refers to the following. If the inflaton is a fundamental scalar, then as usual quantum corrections drive its mass to the cut-off of the effective field theory describing it. This, in particular, implies that $\eta$, one of the ``slow roll" parameters that need to be $<\!\!<1$ during Inflation, is driven to ${\cal O}(1)$.\footnote{The $\eta$-parameter is proportional to the inflaton mass.} As a result, the inflationary expansion ends prematurely \cite{DB}. On the other hand, if the inflaton is a composite scalar in a strongly coupled gauge theory, then its mass is dynamically fixed. In addition, a huge advantage of having a gravitational dual of such composite inflation models is that the inflaton mass becomes a calculable quantity, just like the full glueball mass spectrum is. Finally, it is also worth noting that glueball inflation models are expected to be able to produce large enough levels of primordial gravitational waves \cite{CK}, so that they could be tested within the next years. For all these reasons it is, clearly, of great interest to find gravity duals of glueball inflation. We hope to come back to this issue in the near future.

\section{Discussion} 

We investigated a five-dimensional consistent truncation of type IIB supergravity, relevant for gauge/gravity duality. Our goal was to find solutions with the metric ansatz 
\be
ds_5^2 = e^{2A(z)} g_{\mu \nu} dx^{\mu} dx^{\nu} + dz^2 \,\, ,
\ee
where the 4d metric $g_{\mu \nu}$ is dS or AdS, and the scalars are of the form $\Phi^i = \Phi^i (z)$.  At first sight, this ansatz does not have enough unknown functions to allow for solutions to exist generically. Namely, there is one more field equation than unknown functions. However, we showed that, for a nontrivial warp factor $A(z)$, one of those equations becomes dependent on the others and, thus, is automatically satisfied when they are. This is true regardless of the sign of the 4d cosmological constant $\Lambda$. We also saw that, for $A=const$, there are no solutions with $\Lambda > 0$.

Further, we showed that three of the six scalar fields $\Phi^i=\{p,x,g,\phi,a,b\}$, in the 5d theory under consideration, can be consistently set to zero; see (\ref{gabzero}). That simplification allowed us to find analytically an explicit solution with $\Lambda > 0$, for which:
\be \label{Sol1}
x = - 6p \quad , \quad \phi = 0 \quad , \quad g = 0 \quad , \quad a = 0 \quad , \quad b = 0
\ee
and $p(z)$ and $A(z)$ are given by (\ref{PsolpA}). This solution was obtained in an approximation, similar to the limit in which the walking solutions of \cite{NPP} are found. It would be rather interesting to understand whether there is a conceptual reason for that. Also, the existence of this solution depended crucially on the presence of an overall minus sign in front of the leading term in the {\it five}-dimensional potential. In other words, it was very important that the 5d potential was negative definite, in the appropriate limit. It is definitely worth trying to understand the physical significance of this.

Still working within the truncation (\ref{gabzero}), we were able to find numerically two other classes of solutions with $\Lambda > 0$, for which:
\be \label{Sol2}
x = - \frac{3}{2} p \quad , \quad  \phi = 3p \quad , \quad g = 0 \quad , \quad a = 0 \quad , \quad b = 0 \quad .
\ee
These solutions are obtained, basically, in the limit $P>\!\!>1$, $Q=0$. Furthermore, the coupled system of differential equations for $p(z)$ and $A(z)$ reduces in this case to a single algebraic relation. We were able to find a couple of one-parameter families of numerical solutions to this equation. Interestingly, it turns out that one of these classes of solutions has a finite upper bound, $z_{\rm max}$\,, on the range of $z$. 

The analytical solution in (\ref{Sol1}) and both classes of numerical solutions in (\ref{Sol2}) are regular at the origin $z=0$, for appropriate choice of integration constants. However, only one of the two numerical classes of solutions, the one corresponding to (\ref{z1}), is finite for $z \to \infty$. The other class, as well as the analytical solution, has divergences at the upper bound of $z$ (for the numerical class this is at $z_{\rm max}$\,, whereas for the analytical solution it is at $z\rightarrow \infty$). In physical applications, though, there could be a natural upper cutoff for the radial variable. This could be a dynamically generated scale, above which the effective description provided by this model is not valid anymore, as in \cite{ASW}. Or, in view of the cosmological applications, it could be the Hubble scale, that is the natural effective field theory cutoff \cite{DB}. So our solutions could be useful laboratories for studying strongly-coupled gauge theories in de Sitter space, regardless of whether they have large $z$ divergences or not. Of course, that will also depend on whether or not there are perturbative instabilities. Investigating the latter issue is rather involved and we leave it for the future. Another interesting question, worth addressing, is what microscopic (D-brane) realizations could lead to our solutions.

Finally, the explicit analytical solution in (\ref{Sol1}) could provide a useful starting point for finding a gravitational dual of Glueball Inflation. Indeed, since in slow roll Inflation the Hubble parameter has to vary with time rather slowly, the inflating solution can be viewed as a small time-dependent perturbation of pure dS space (which has constant Hubble parameter). Hence, studying small time-dependent perturbations around (\ref{Sol1}), together with (\ref{PsolpA}), could enable one to find an explicit solution with an inflating 4d spacetime. We intend to explore this problem in the near future.

\section*{Acknowledgements}

We would like to thank C. Nunez, M. Piai, G. Tasinato and I. Zavala for discussions on gravitational duals of strongly coupled gauge theories and on cosmological inflation. L.A. is grateful to the Simons Workshop in Mathematics and Physics, Stony Brook 2014, for hospitality during the completion of this work. L.A. has received partial support from COST Action MP-1210 and Bulgarian NSF grant DFNI T02/6 during the work on this project.

\appendix

\section{Parameter range for large $P$ solutions}
\setcounter{equation}{0}

In Section 4.2.2 we found numerical solutions to equation (\ref{T}), which have the form (\ref{Tser}) in a neighborhood of $w=0$. Using these, one can find numerically the functions of interest, namely $q(z)$ and $B(z)$. However, it turns out that not every value of the integration constant $t_1$, that parametrizes this family of solutions, leads to real $B(z)$. Recall that, to give a physical solution via (\ref{sub1}), the function $B(z)$ has to be real and positive. We will show now that only $t_1 \in \left(-\infty, - \frac{16}{7}\right) \cup (0,\infty)$ leads to physical solutions for small $T$.

By inverting (\ref{Tser}), we can obtain the expansion of $q$ in powers of $T$. Substituting this expression into (\ref{Bsol}), we find the following expansion for $B^2$:
\be \label{Bser}
B^2 = \frac{7}{144} \,t_1 \,(16+7 t_1) + \left( \frac{7}{24}+\frac{2}{9 t_1}+\frac{49 t_1}{576} \right) T + ... \,\,\, .
\ee
Recall that the constant $t_1 \neq 0$, as discussed above (\ref{Tser}). So (\ref{Bser}) is well-defined. Furthermore, for $T \to 0$, the first term determines the sign of $B^2$. Hence, we see that for small $T$ we have $B^2 > 0$ (and thus real nonzero $B$) for $t_1 < - \frac{16}{7}$ or $t_1 > 0$\,, while $B^2<0$ for $t_1 \in \left(- \frac{16}{7} , 0\right)$. Also, one can see that $B=0$ for $t_1 = - \frac{16}{7}$, as the coefficient of each term in the $B^2 (T)$ expansion is zero. Note that an identically vanishing $B$ is not a physical solution, since it gives an infinite warp factor $A$ for any $z$; see (\ref{sub1}). Hence, we conclude that the parameter range $t_1 \in \left[- \frac{16}{7} , 0\right]$ is not allowed for physical solutions following from (\ref{Tser}).

As an illustration of the above, let us briefly discuss the special case corresponding to $t_1 = - \frac{16}{7}$. Interestingly, one can find this solution by solving 
\be\label{Teq}
T' = -2\,T-\frac{8}{7}\pm\frac{4}{7}\,\sqrt{4+11\,T+7\,T^2} \,\, .
\ee
It is easy to verify that any $T$, satisfying (\ref{Teq}), automatically satisfies (\ref{T}) as well, although the reverse is not true. The solution with $t_1 = - \frac{16}{7}$ is obtained for the minus sign in (\ref{Teq}), whereas the plus sign gives a solution with diverging $q$ for small $T$. Indeed, imposing the constraint $T(0)=0$, one can easily verify that the Taylor expansion coefficients, for the solution of the minus-sign equation in (\ref{Teq}), coincide with those in (\ref{Tser}) with $t_1 = - \frac{16}{7}$.

We will show now that $B \equiv 0$ for any solution of (\ref{Teq}). By integrating (\ref{Teq}), we find: 
\be\label{Tsoln}
\frac{q}{q_0}=\left[\frac{3}{2\,T^2}\,(8+11\,T\pm 4\,\sqrt{4+11\,T+7\,T^2})\right]^{\frac{7}{6}}\!(11+14\,T\pm 2\,\sqrt{7}\,\sqrt{4+11\,T+7\,T^2})^{-\frac{\sqrt{7}}{3}},
\ee
where $q_0 = e^{w_0}$. This gives us $q$ as a function of $T$. To compute $B$ from (\ref{Bsol}), we also need $q'(z)$ and $q''(z)$ as functions of $T$. From (\ref{betaeq}), we have that $q'$ is given by:
\be\label{qp}
q'(z)= \pm \frac{7}{3} \,q(T)^2 \sqrt{1+T} \,\, .
\ee
Then, taking the derivative of (\ref{qp}) with respect to $z$ and substituting (\ref{qp}) in the result, we obtain:
\be\label{qpp}
q''(z)=\pm  q'(z) \,q(T)\,\frac{7}{6\,\sqrt{1+T}}\left[4\,(1+T)+ q\,\frac{dT}{dq}\right]=\frac{49}{18}\,q(T)^3\left[4(1+T)+\frac{q(T)}{q'(T)}\right],
\ee
where $q(T)\,/\,q'(T)$ can be calculated by differentiating (\ref{Tsoln}) with respect to $T$.

Substituting into (\ref{Bsol}) the expressions for $q$, $q'$ and $q''$ from (\ref{Tsoln}), (\ref{qp}) and (\ref{qpp}) respectively, we obtain $B=0$ regardless of the choice of sign in (\ref{Tsoln}). Clearly then, these solutions are physically unacceptable, as they lead to infinite warp factor $A$ for any $z$, according to (\ref{sub1}).

\end{document}